\documentclass[journal,transmag]{IEEEtran}

\usepackage{graphicx}
\usepackage{caption,cite}
\usepackage{subcaption}
\usepackage{multirow}
\usepackage{gensymb}
\usepackage{amsmath}
\usepackage{lipsum}
\usepackage{color}
\usepackage{array}
\newcolumntype{P}[1]{>{\centering\arraybackslash}p{#1}}

\begin{document}

\title{Comparison Between Different Designs and Realizations of Anomalous Reflectors}

\author{\IEEEauthorblockN{Mostafa Movahediqomi\IEEEauthorrefmark{1},
Grigorii Ptitcyn\IEEEauthorrefmark{1}, and
Sergei Tretyakov\IEEEauthorrefmark{1},~\IEEEmembership{Fellow,~IEEE}}
\IEEEauthorblockA{\IEEEauthorrefmark{1}Department of Electronics and Nanoengineering, School of Electrical Engineering, Aalto University, 02150 Espoo, Finland}
}

\IEEEtitleabstractindextext{
\begin{abstract}
Metasurfaces enable efficient manipulation of electromagnetic radiation. In particular, control over plane-wave reflection is one of the most useful features in many applications. Extensive research has been done in the field of anomalous reflectors over the past years,  resulting in numerous introduced geometries and several distinct design approaches. Anomalously reflecting metasurfaces designed using different methods show different performances in terms of reflection efficiency, angular response, frequency bandwidth, etc.  Without a  comprehensive comparison between known design approaches, it is difficult to properly select the most appropriate design method and the most suitable metasurface geometry.  Here, we consider four main approaches that can be used to design anomalous reflectors within the same basic topology of the structure and study the designed metasurfaces first on the level of the input impedance and then consider and compare the performance of the realized structures. We cover a wide range of performance aspects, such as the power efficiency and losses, angular response, and the scattering pattern of finite-size structures. 
We anticipate that this study will prove useful for developing new engineering methods and designing more sophisticated structures that include reconfigurable elements. Furthermore, we believe that this study can be considered referential since it provides comparative physical insight into anomalous reflectors in general.

\end{abstract}

}

\maketitle

\IEEEdisplaynontitleabstractindextext

\IEEEpeerreviewmaketitle

\section{Introduction}
\label{Section1}

Wireless communication technologies constantly progress towards higher operational frequencies. This progress comes with smaller antenna sizes and, alas, at the expense of the need to use highly-directive and scanning antennas. Improvement of transmitters and receivers is limited, therefore communication engineers proposed to optimize the propagation environment using metasurfaces and metagratings ~\cite{asadchy2016perfect,estakhri2016wave,epstein2016synthesis,ra2017metagratings,budhu2020perfectly,kwon2018lossless,wang2020independent,kwon2021planar,yepes2021perfect,diaz2017generalized,wong2018perfect,rabinovich2018analytical,rabinovich2019experimental,diaz2019power}, and reconfigurable intelligent metasurfaces (RIS)~\cite{basar2019wireless,liu2019intelligent,di2020smart,di2020analytical,pitilakis2020multi,tsilipakos2020toward,di2020reconfigurable,degli2022reradiation,di2021communication,danufane2021path}. The latter approach has gained increasing attention recently in communication communities. Often, reconfigurable structures are designed based on conventional fixed structures with the addition of tunable elements. Therefore, a comparison of known approaches to design anomalous reflectors is timely.

There are two fundamentally different methods to realize a flat surface that reflects plane waves into plane waves along any desired direction. One is the use of periodical structures (diffraction gratings) whose period is chosen accordingly to the required angles of incidence and reflection. The other one is using aperiodically loaded antenna arrays whose geometrical period is fixed to usually $\lambda/2$ \cite{Fu_tutorial}. The majority of works on anomalous reflectors use the first approach, and here we consider various designs of periodically modulated anomalously reflected boundaries.  

Perhaps, the most classical approach to manipulate the direction of reflection from a surface is the use of phased-array (reflectarray) antennas~\cite{berry1963reflectarray,huang2007reflectarray,nayeri2018reflectarray}. Here, the phase distribution at the antenna aperture is tuned so that reflections from all antenna array elements interfere constructively along the desired direction of reflection. Generalizing this principle, a similar approach can be realized in a planar subwavelength-structured metasurface if the local reflection phase is made nonuniform over the surface, realizing a phase-gradient reflector. Using this approach one can direct the reflected wave at will, beating the conventional law of reflection and realizing so-called anomalous  reflection. The main drawback of this method is the low efficiency at large deviations from the usual law of reflection~\cite{estakhri2016wave, asadchy2017eliminating}.
Impedance mismatch between the incident and the reflected waves becomes significant, and it causes more scattering into parasitic propagating modes (See Fig.~\ref{fig:conceptual}).

Theoretically, the problem of reduced efficiency at large deflection angles can be completely solved with the use of active and lossy inclusions in the metasurface~\cite{asadchy2016perfect}. Worth mentioning, that the average power produced by the surface would be zero, however, some parts of it must produce energy, and the other parts should absorb it, which is quite impractical. Another possibility is to use completely passive structures, where auxiliary surface waves in the near-field region are properly tuned~\cite{asadchy2016perfect,epstein2016synthesis} as it is shown in Fig.~\ref{fig:conceptual}. Optimization of the evanescent modes can be performed in several different ways: based on the optimization of the input (surface) impedance~\cite{budhu2020perfectly, kwon2018lossless}, grid (sheet) impedance~\cite{wang2020independent, yepes2021perfect, kwon2021planar},  by direct optimization of the whole structure~\cite{diaz2017generalized, wong2018perfect}, by finding an analytical solution~\cite{rabinovich2018analytical, rabinovich2019experimental}, and finally, introducing non-planar (power flow-conformal) structures~\cite{diaz2019power}. In this paper, we overview and compare some of these approaches in detail and discuss their differences and advantages. We repeated the selected design methods and compared the most important characteristics of these works, including  power efficiency, angular stability, far-field radiation patterns, and frequency bandwidth for the infinite and finite-size structures.

The paper is organized as follows: In Sec.~\ref{Section2} the selected methods for comparison will be briefly introduced and the pros and cons for each of them will be highlighted. Then~Sec.~\ref{section3} is devoted to the investigation of scattering parameters for both ideal and realized structures and the comparison of the power efficiency of each method. The angular response of an anomalous reflector is another important aspect that is discussed  in~Sec.~\ref{section4}. Here, we show the behavior of reflectors when they are illuminated by waves at different angles. Furthermore, reflection and scattering by a finite-size structure in the far zone is important for applications, and  recently it has been considered in several studies.  We cover this issue in~Sec.~\ref{section5}. Finally, conclusions are formulated  in~Sec.~\ref{section6} to finalize this comparison and make the advantages and drawbacks of each approach clear.

\section{Considered design methods}
\label{Section2}

To provide a fair comparison, we choose methods that  can be realized using arrays of metallic patches or strips printed on a grounded dielectric substrate. We select an example required performance: an anomalous reflection of normally incident plane waves with TE polarization to the $70^\circ$-direction, at $8$~GHz.  All designs are based on the same basic platform: a metal patch array on a grounded dielectric substrate  (Fig.~\ref{fig:conceptual}). The chosen example substrate is Rogers 5880 with  2.2 permittivity, 1.575~mm thickness, and 0.0002 loss tangent. For all designs, we split the period into 6 sub-cells that are either impedance strips or shaped metal strips. The use of the same parameters for all designs allows a meaningful comparison of performance.

The initial reference design is a phase-gradient metasurface, e.g.  \cite{berry1963reflectarray,huang2007reflectarray,nayeri2018reflectarray,yu2011light}.  The unit cells are designed in the conventional locally periodical approximation so that at every point of the reflector the reflection phase (at normal incidence) from an infinite array of identical cells is as required by the linear phase gradient rule for the desired reflection angle. 
It means that the reflection properties of a metasurface can be defined by the ``local reflection coefficient'' which is assumed to be controlled by adjusting the geometrical parameters of the unit cells. Strong coupling between the inclusions in an inhomogeneous array  makes this approximation rather rough when the deflection angle is not small. 

More advanced methods that aim at overcoming the inherent parasitic scattering of phase-gradient reflectors we classify based on the degree of use of homogenized boundary conditions:
\begin{figure}[t!]
     \centering
         \includegraphics[width=0.5\textwidth]{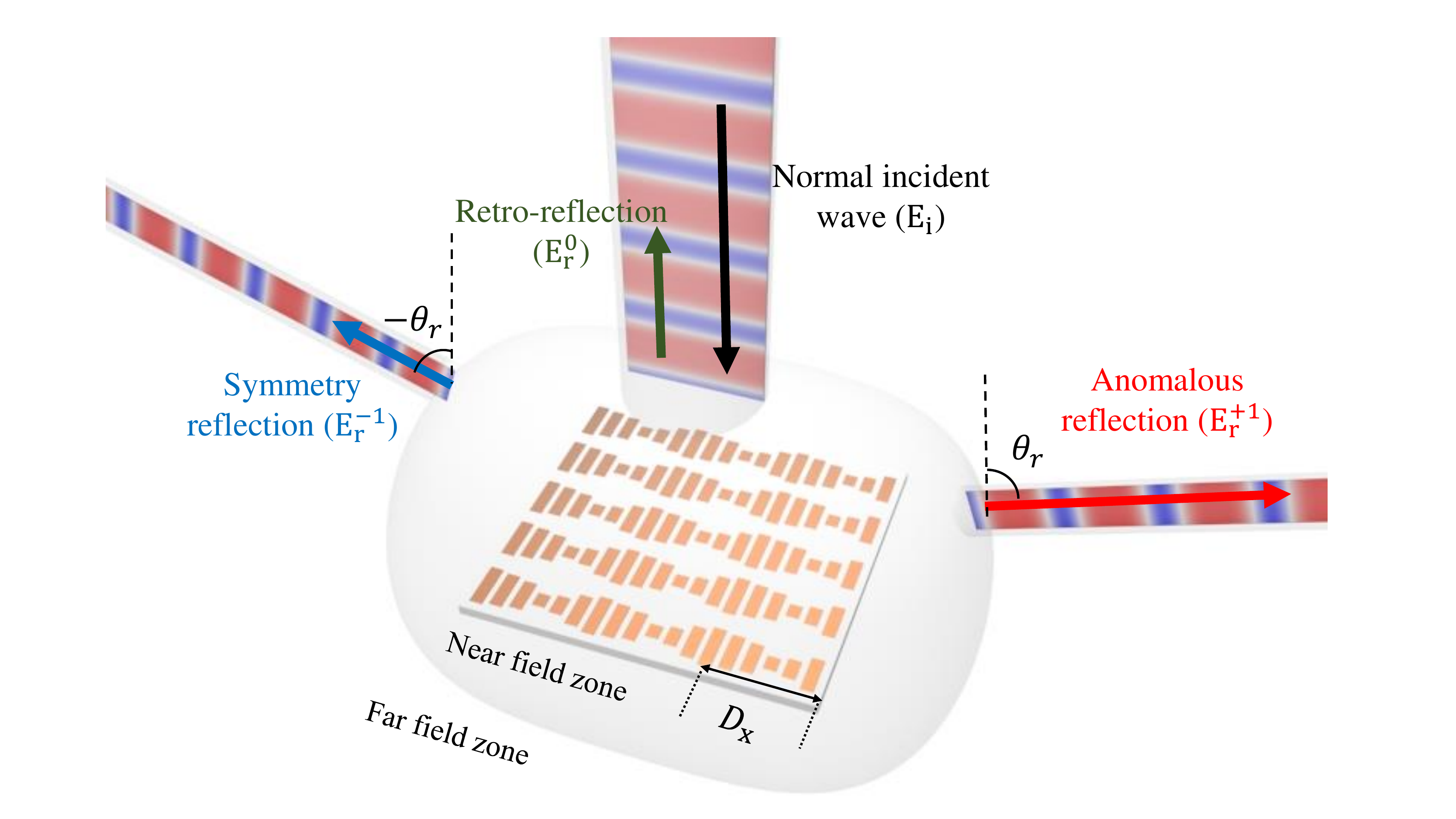}
    \caption{Concept of periodical arrays acting as anomalous reflectors. The case with three propagating Floquet harmonics is illustrated. The design goal is to suppress reflections into all propagating modes except the desired anomalous reflection. }
    \label{fig:conceptual}
\end{figure}

\begin{figure}[t!]
     \centering
        \includegraphics[width=0.5\textwidth]{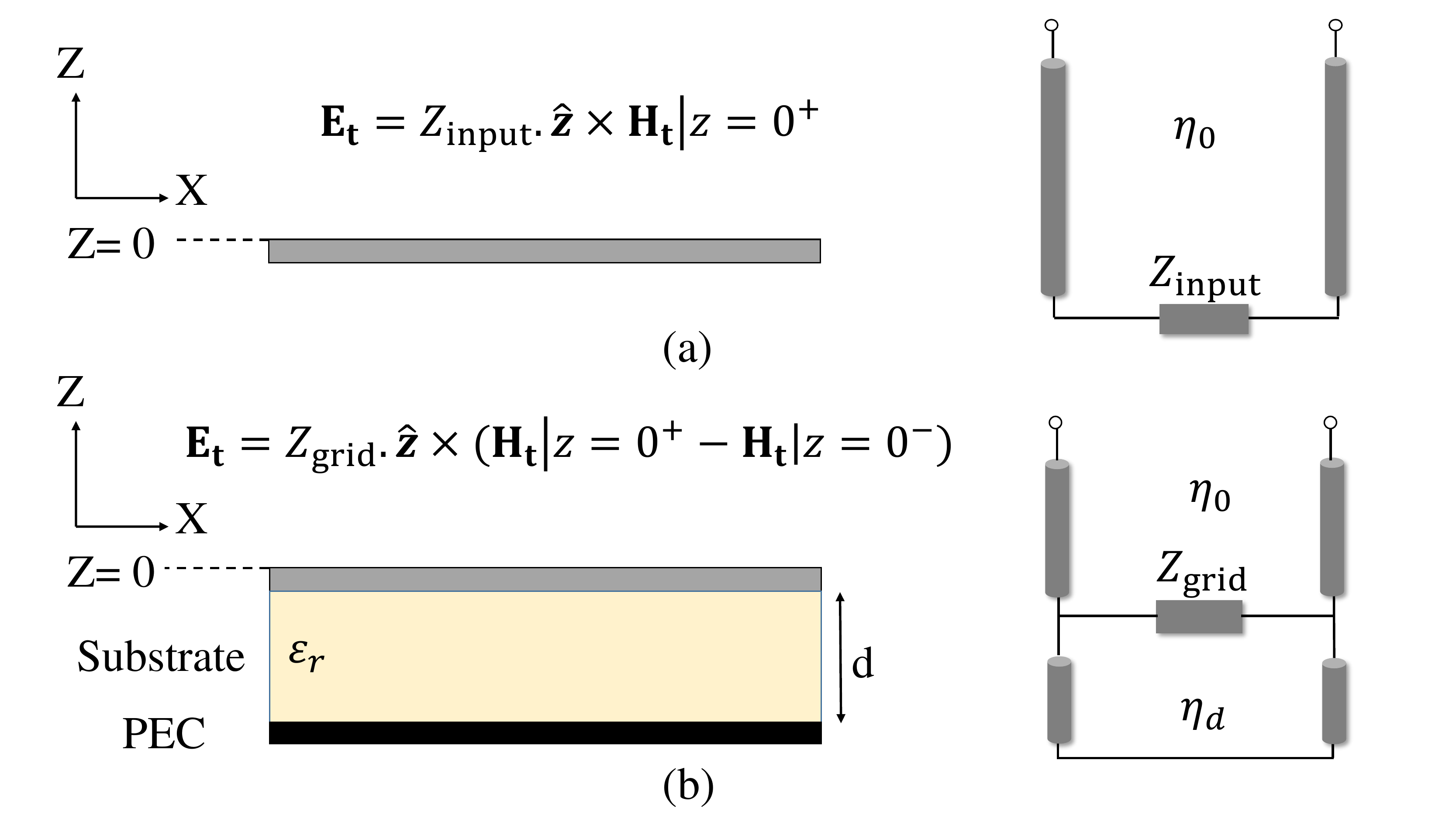}
        \caption{Two types of IBCs: (a) input impedance, also known as impenetrable IBC. The left side illustrates the conceptual structure, and the right side shows the corresponding transmission-line model. (b) Grid or sheet impedance is also known as penetrable IBC. The conceptual structure on the left consists of an impedance sheet placed on top of a grounded dielectric substrate. The equivalent transmission-line model is shown on the right side.}
        \label{fig:IBCs}
\end{figure}

Method~1 (input impedance method).  Here, the metasurface is designed at the level of the equivalent input impedance, also known as the impenetrable impedance boundary condition (IBC), see Fig.~\ref{fig:IBCs}(a). The input impedance ($Z_{\rm input}$) relates the tangential components of the electric field ($\boldsymbol{\rm E}_{\rm t}$) and magnetic field at the interface between the metasurface structure and free space: ($\boldsymbol{\rm H}_{\rm t}$)
\begin{equation}
\label{Eq:input_imp}
\boldsymbol{\rm E}_{\rm t}=Z_{\rm input}\cdot\hat{\boldsymbol{z}}\times\boldsymbol{\rm H}_{\rm t}\mid_{z=0^+}.
\end{equation}
In this method, the input impedance distribution over the reflector surface is optimized with the goal to  channel most of the reflected power into the desired direction. Optimization algorithms vary the input impedance, ensuring zero normal component of the Poynting vector at every point of the surface so that the input impedance is purely reactive~\cite{budhu2020perfectly, kwon2018lossless}. When the desired input impedance values at every point are found, the actual geometry of the structure is determined using the locally periodic approximation. That is, the continuous reactance profile is discretized, and the dimensions of each unit cell are optimized using periodical boundary conditions, ensuring that the plane-wave reflection phase (at normal incidence) from an infinite periodical array of this cell is the same as from a uniform boundary with the required input reactance at this point.

Method 2 (grid impedance method). This design approach is based on the grid (or sheet) impedance model of a patch array that is also known as penetrable IBC, see Fig.~\ref{fig:IBCs}(b). In this method, the impedance boundary condition is used to model only the array of metal patches. The grid impedance ($Z_{\rm grid}$) relates the surface-averaged electric field with the difference between the averaged tangential magnetic fields at both sides of the metasheet:
\begin{equation}
\label{Eq:grid_imp}
\boldsymbol{\rm E}_{\rm t}=Z_{\rm grid}\cdot\hat{\boldsymbol{z}}\times(\boldsymbol{\rm H}_{\rm t}\mid_{z=0^+}-\boldsymbol{\rm H}_{\rm t}\mid_{z=0^-}).
\end{equation}
In this method, spatial dispersion of the grounded dielectric layer is taken into account. 
The optimization process in this case considers a more practical structure, that treats waves inside the substrate in a more complete way compared to the first method~\cite{wang2020independent, kwon2021planar, yepes2021perfect}. 
In method 2 the locally periodical approximation is used to design reactive sheets in contrast with method 1 in which it is utilized to model the whole metasurface volume.

\begin{figure*}
    \centering
         \includegraphics[width=1\textwidth]{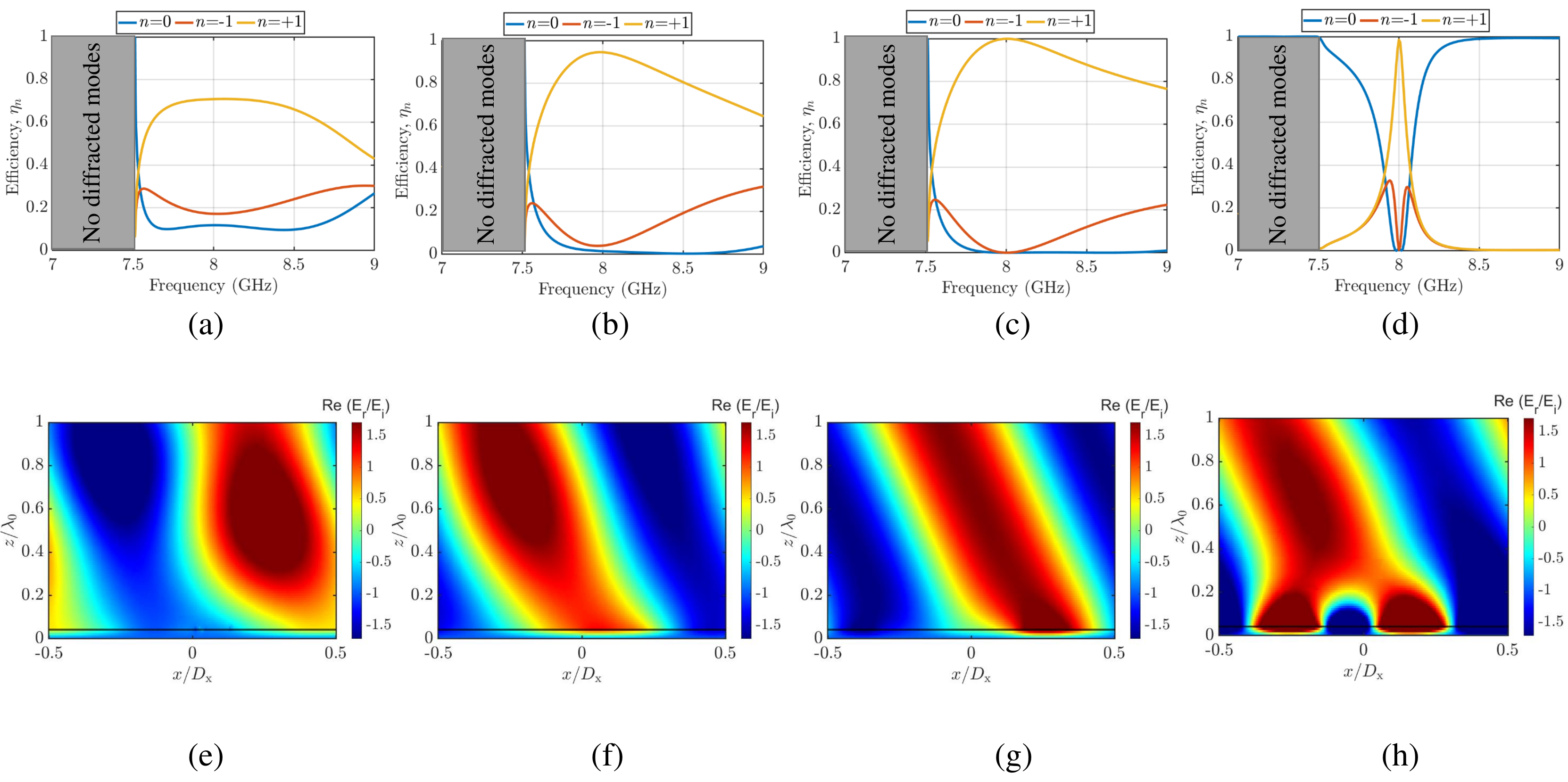}
    \caption{Power distribution between three  propagating modes and scattered field distribution (bottom); (a,e) for the phase gradient, (b,f) input impedance, (c,g) grid impedance, (d,h) non-local design method. The horizontal black lines in the scattered field distribution figures illustrate the location of the metasurfaces where the IBC is applied.}
    \label{fig:scattering_par_field}
\end{figure*}

Method 3 (non-local design) accounts for all specific geometrical and electromagnetic features of the layer, not relying on homogenization methods. The optimization usually starts from some initial settings in terms of the input impedance (for example, in \cite{diaz2017generalized} it was required that the reflector if formed by  periodically arranged regions of receiving and re-radiating leaky-wave antennas), but the final steps optimize the whole supercell of the periodical lattice instead of individual patches in periodical arrays. Importantly, the normal component of the Poynting vector along the surface is not set to zero, corresponding to the effective active-lossy behavior, although the overall structure remains completely passive. The drawback of this approach lies in the need for direct optimization, which usually requires heavy computational facilities and might also become time-consuming. 

Other approaches realize perfect anomalous reflection using arrays of loaded wires \cite{rabinovich2018analytical,rabinovich2019experimental} or  non-planar structures~\cite{diaz2019power}. However, to provide an insightful comparison, we chose only  methods that are suitable for planar  structures that can be realized as printed circuit boards with metallic patches. Specifically,  the phase-gradient sample is designed based on the required tangent-profile of the input impedance, for method~1 (impenetrable IBCs) we follow~\cite{kwon2018lossless}, paper~\cite{wang2020independent} for method~2 (penetrable IBCs), and~\cite{diaz2017generalized} for method~3 (non-local design). 
In the following sections, the scattering properties, angular response, as well as far-field characteristics of test finite-size structures for all aforementioned methods will be investigated and compared in detail.

\section{Scattering properties}
\label{section3}
At first, the scattering properties of all the anomalous reflectors under study will be investigated assuming infinite periodical structures. Upon plane-wave illuminations, the structures  support surface currents that are also periodic. The Floquet theory defines the tangential wavenumbers of modes supported by the surface:
\begin{equation}
\label{Eq:tan_wavenumber}
k_{\rm t}=k_{\rm t0}+k_{\mathrm{t}n}=k_{0}\sin{\theta_{\rm i}+2\pi} n/D,
\end{equation}
Where $k_0$ is the wavenumber in free space, $\theta_{\rm i}$ is the angle of the upcoming incident wave, $n$ is an integer number that denotes the index of the Floquet mode, and $D$ is the period of the surface pattern,  determined by $D=\lambda/ (\sin{\theta_{\rm r}}-\sin{\theta_{\rm i}})$. $\theta_{\rm r}$ is the desired reflection angle. By choosing this period, the tangential component of the wavenumber is fixed so that one of the harmonics is reflected to the desired angle. Floquet harmonics that satisfy criterion $k_0>|k_t|$ belong to the fast-wave regime in the dispersion diagram and can propagate in free space. Other Floquet harmonics are surface waves. The direction of the reflection can be calculated by the following formula:
\begin{equation}
\label{Eq:wave_direction}
\sin\theta_{\rm r}=k_{\rm t}/k_0=(k_{0}\sin{\theta_{\rm i}+2\pi} n/D)/k_0.
\end{equation}
For the chosen design parameters ($\theta_{\rm i}=0\degree$, $\theta_{\rm r}=70\degree$, $f=8$~GHz), the period is equal to $D=1.0642\lambda$, and the Floquet expansion has three propagating harmonics ($k_0>|k_t|$): zero Floquet mode (0\degree), $-1$ Floquet mode ($-70\degree$), and $+1$ Floquet mode ($+70$\degree), as follows from  Eq.~\eqref{Eq:wave_direction}. The field amplitudes in these  modes define the efficiency of power channeling from one mode to another.

\subsubsection{Performance of the surface-impedance models}

Initially, the ideal impedance profile is considered when the period is discretized to six elements. In other words, it is assumed that the impedance boundary condition is applied straight on the surface without considering actual realization (Fig.~\ref{fig:IBCs}). It is noteworthy to notice that the discretization of the impedance profile deteriorates performance, however, it is spoiled in the same way for all methods. Using such discretization,  reasonable results can be achieved rather fast. All the methods except the phase gradient method use optimization, therefore analytical closed-form formulas for the impedance profiles do not exist. The list of optimized impedance values of each unit cell in a period is presented in Table~\ref{Table:impedance values} for all designed methods. For both designs based on the input impedance model (phase-gradient and input impedance optimization), we convert the obtained input impedance profile to the grid impedance by using the equivalent transmission-line model presented  in~Fig.~\ref{fig:IBCs}. The input impedance can be considered as that of a shunt connection of the grid impedance to the transmission line modeling the grounded dielectric substrate. For the non-local method, a pre-final optimization is applied here similarly to what was done in~\cite{diaz2017generalized}. As it was discussed in~Sec.~\ref{Section2}, for the non-local approach we can assume an impedance profile using repeated receiving and re-radiating leaky-wave sections to mimic the ideal active-lossy profile. Therefore, the optimization at the level of the grid impedance is an initial step before the final optimization for the whole supercell in the real structure. Eventually, the same configuration for all
the methods enables us to complete a fair comparison.

\begin{table}[!h]
\renewcommand{\arraystretch}{1.3}
\caption{The impedance profile list for each unit cell ($j\Omega$)}
\label{Table:impedance values}
\centering
\begin{tabular}{p{0.055\textwidth}p{0.045\textwidth}p{0.049\textwidth}p{0.045\textwidth}p{0.045\textwidth}p{0.045\textwidth}p{0.045\textwidth}}
\hline
 & \bfseries Cell1 & \bfseries Cell2 & \bfseries Cell3 & \bfseries Cell4 & \bfseries Cell5 & \bfseries Cell6\\
\hline\hline
Phase gradient & -97.6 & -82.1 & -51.4 & +2673.9 & -145.4 & -113.4\\
\hline
Input impedance & -291.7 & -141.3 & -114.0 & -98.0 & -83.8 & -85.3\\
\hline
Grid impedance & -73.8 & -1334.0 & -112.9 & -172.6 & -91.4 & -96.7\\
\hline
Non-local & -81.82 & -74.74 & -49.16 & -73.50 & -75.07 & -73.06\\
\hline
\end{tabular}
\end{table}

Performance comparison of the discretized impedance profiles after optimization is made using full-wave simulators, CST STUDIO~\cite{CST} and ANSYS HFSS~\cite{HFSS}. As it was discussed, there are three propagating Floquet harmonics (open channels) in our specific example. Therefore, we can consider these reflectors as three-port networks. Scattering parameters ($S_{n1}$) can be determined numerically when the input wave comes from Floquet port $1$ and the output wave is observed in the port number $n$. Consequently, the power efficiency is found as squared scattering parameters ($\eta_n=|S_{n1}|^2$) in the full-wave simulators. The power efficiency for each mode measures the fraction of power rerouted from the incident wave (assuming that the incident port is $1$) to the propagating mode $n$.

Figures~\ref{fig:scattering_par_field} (a-d) show ratios of power rerouted to propagating channels $n=0, 1$,  and $-1$. In all cases, below $7.5$~GHz diffraction modes are not allowed, therefore all the energy is reflected back to the normal direction. Designs based on the phase-gradient, input impedance, and grid impedance methods show broadband behavior as compared to the non-local approach. The phase-gradient method does not take evanescent modes into account, which results in the lowest efficiency at the operational frequency. Power distribution and the corresponding field amplitudes for all methods can be found in Table~\ref{Table:amp and eff ideal imp}.
 
\begin{table}[!h]
\renewcommand{\arraystretch}{1.3}
\caption{Amplitude/power ratio of propagating Floquet modes and power efficiency level at $8$~GHz}
\label{Table:amp and eff ideal imp}
\centering
\begin{tabular}{p{0.15\textwidth}p{0.07\textwidth}p{0.07\textwidth}p{0.07\textwidth}p{0.07\textwidth}}
\hline
 & \bfseries  $\theta_{\mathrm{i}}$ & \bfseries $\theta_{\mathrm{r}}$ & \bfseries $-\theta_{\mathrm{r}}$ \\
\hline\hline
Phase gradient & 0.33/0.11 & 1.45/0.72 & 0.71/0.17\\
\hline
Input impedance & 0.10/0.01 & 1.67/0.95 & 0.34/0.04\\
\hline
Grid impedance & 0.00/0.00 & 1.71/1 & 0.01/0.00\\
\hline
Non-local & 0.1/0.01 & 1.70/0.99 & 0.12/0.00\\
\hline
\end{tabular}
\end{table}

It is noteworthy to sketch the scattered electric field distributions (Fig.~\ref{fig:scattering_par_field}(e-h)). 
Efficiency for the phase-gradient method is only 71.8$\%$, and, correspondingly, Fig.~\ref{fig:scattering_par_field}(e) shows a  field distribution that is distorted by fields scattered into two parasitic propagating channels. For the other methods, efficiency is nearly perfect, however, the near-field distributions are different due to different methods used to optimize evanescent modes.
It is important to note that for perfect anomalous reflection with ideal power efficiency, the power reflected to the desired direction must be equal to the power of the incident plane wave, and, as a result, the ratio between the amplitudes of the reflected and incident fields for these angles should be larger than one  $|E_{\mathrm{r}}|=|E_{\mathrm{i}}|\sqrt{\cos(\theta_{\mathrm{i}})/\cos(\theta_{\mathrm{r}})} $ (1.71 for our example case) \cite{diaz2017generalized}.

\subsubsection{Realizations as patch arrays}

\begin{figure}[ht]
     \centering
         \includegraphics[width=0.45\textwidth]{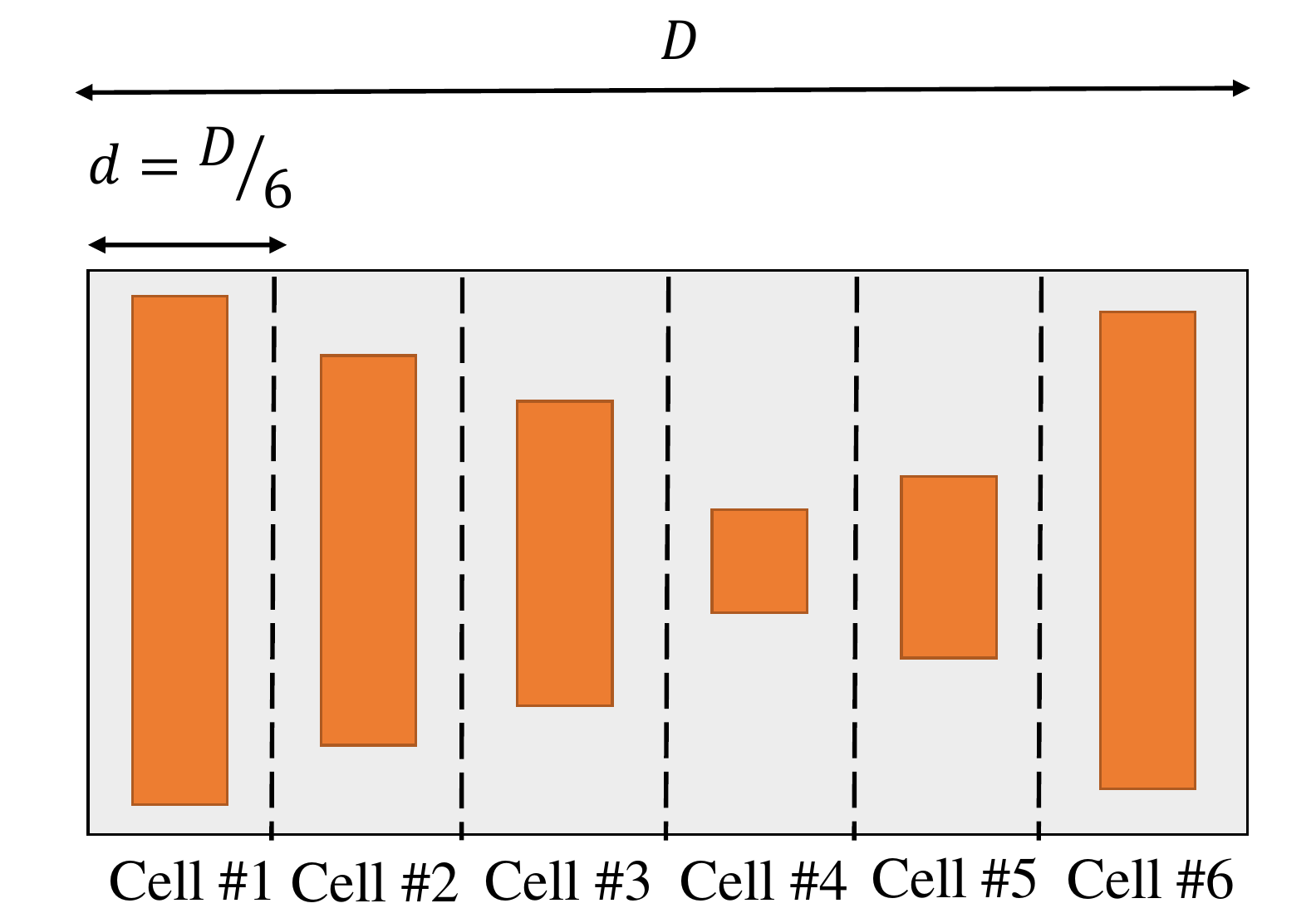}
        \caption{The configuration of supercells utilized for the designs consisting of six unit cells. All the parameters of the dielectric substrate are given  in~Sec.~\ref{Section2}. The period of the array (the supercell size) is fixed to $D=39.9$~mm, and the width of a single unit cell is $d=D/6$. The width of metal strips is $3.5$~mm, while the strip lengths are different for different design methods.}
        \label{fig:supercell}
\end{figure}

The next step is to compare actual structures designed using the  previously obtained and discussed impedance profiles.  Following the  procedures described in the corresponding papers, we design supercells formed by six unit cells based on the rectangular shape metal patches above the grounded dielectric substrate (see~Fig.~\ref{fig:supercell} and Table~\ref{table:length of strips values}).

\begin{figure}[ht]
     \centering
         \includegraphics[width=0.45\textwidth]{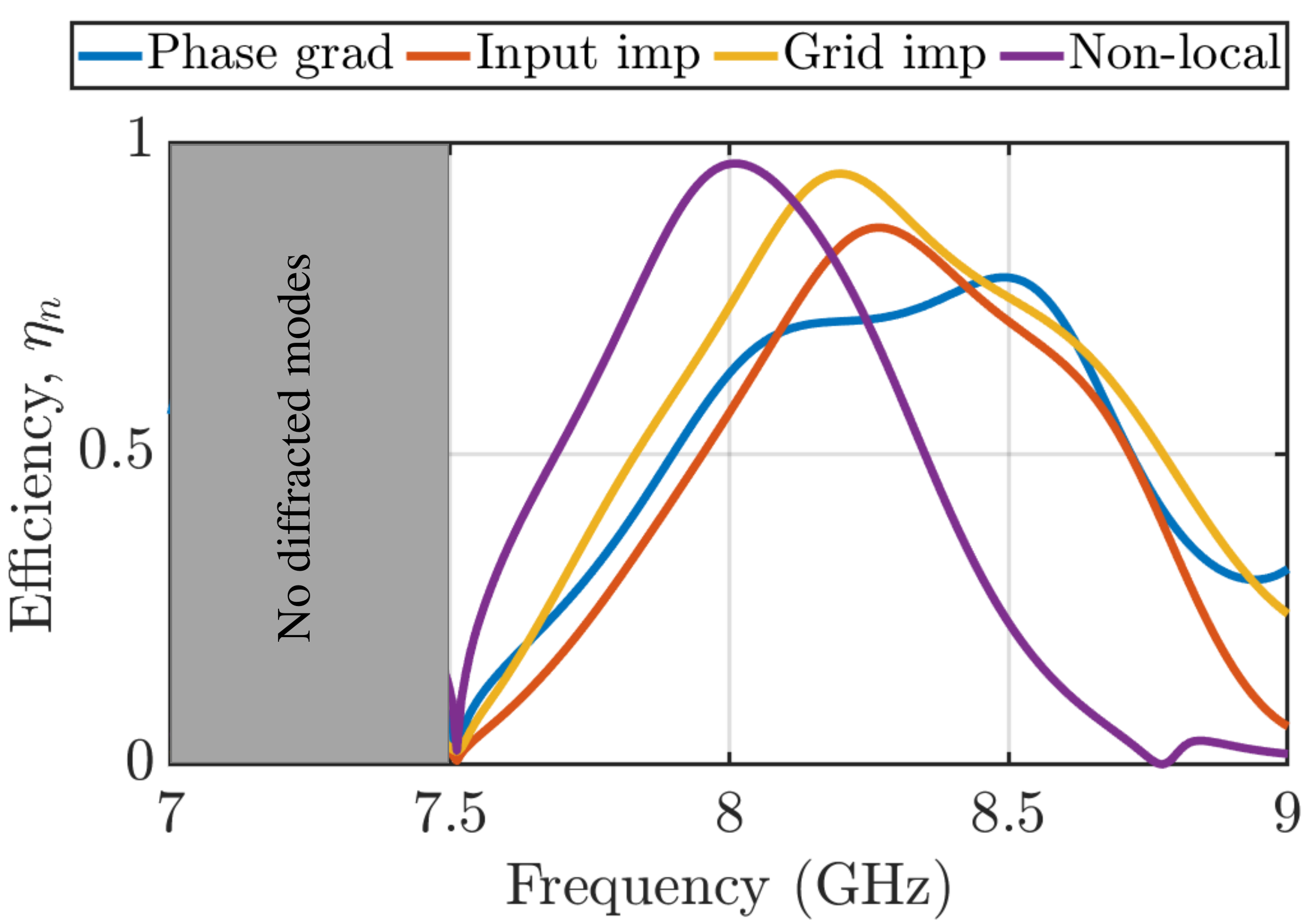}
        \caption{Frequency dependence of efficiency  for structures realized with metallic rectangular patches.}
        \label{fig:metal_strips}
\end{figure}

The corresponding efficiencies for all the considered methods are shown in Fig.~\ref{fig:metal_strips}. The frequency for the best performance becomes shifted for all methods, except for the  non-local design, where optimization of the whole supercell is implemented. In addition to that, dispersion and losses deteriorate the efficiency in different ways. The absorption levels as well as efficiency at the design frequency ($8$~GHz) are reported in Table~\ref{table:realized structure}. The remained power is scattered to other propagating Floquet modes that are not shown in Fig.~\ref{fig:metal_strips}. 

\begin{table}[!h]
\renewcommand{\arraystretch}{1.3}
\caption{Lengths of metal strips for each unit cell (mm)}
\label{table:length of strips values}
\centering
\begin{tabular}{p{0.06\textwidth}p{0.045\textwidth}p{0.055\textwidth}p{0.045\textwidth}p{0.045\textwidth}p{0.045\textwidth}p{0.045\textwidth}}
\hline
 & \bfseries Strip1 & \bfseries Strip2 & \bfseries Strip3 & \bfseries Strip4 & \bfseries Strip5 & \bfseries Strip6\\
\hline\hline
Phase gradient & 10.8 & 11.41 & 13.23 & 0 & 9.47 & 10.29\\
\hline
Input impedance & 7.3 & 9.57 & 10.28 & 10.78 & 11.3 & 11.25\\
\hline
Grid impedance & 3.71 & 10.32 & 8.89 & 11.03 & 10.84 & 11.78\\
\hline
Non-local & 10.47 & 10.91 & 11.26 & 12.22 & 11.30 & 8.88\\
\hline
\end{tabular}
\end{table}

\begin{table}[!h]
\renewcommand{\arraystretch}{1.3}
\caption{The best-performance frequency and the corresponding efficiency versus the absorption rate and efficiency for the design frequency.}
\label{table:realized structure}
\centering
\begin{tabular}{p{0.09\textwidth}p{0.070\textwidth}p{0.070\textwidth}p{0.070\textwidth}p{0.070\textwidth}}
\hline
 &  \multicolumn{2}{c}{Best performance} & \multicolumn{2}{c}{$8$~GHz}\\
\hline
 & \bfseries Frequency & \bfseries Efficiency & \bfseries Absorption & \bfseries Efficiency\\
\hline\hline
Phase gradient & $8.5$~GHz & 78.2($\%$) & 3.9($\%$) & 62.7($\%$)\\
\hline
Input impedance & $8.27$~GHz & 86.3($\%$) & 3.2($\%$) & 56.6($\%$)\\
\hline
Grid impedance & $8.2$~GHz & 95.0($\%$) & 3.3($\%$) & 73.6($\%$)\\
\hline
Non-local & $8$~GHz & 96.6($\%$) & 3.5($\%$) & 96.6($\%$)\\
\hline
\end{tabular}
\end{table}

\section{Angular response}
\label{section4}
\begin{figure}[ht]
     \centering
         \includegraphics[width=0.45\textwidth]{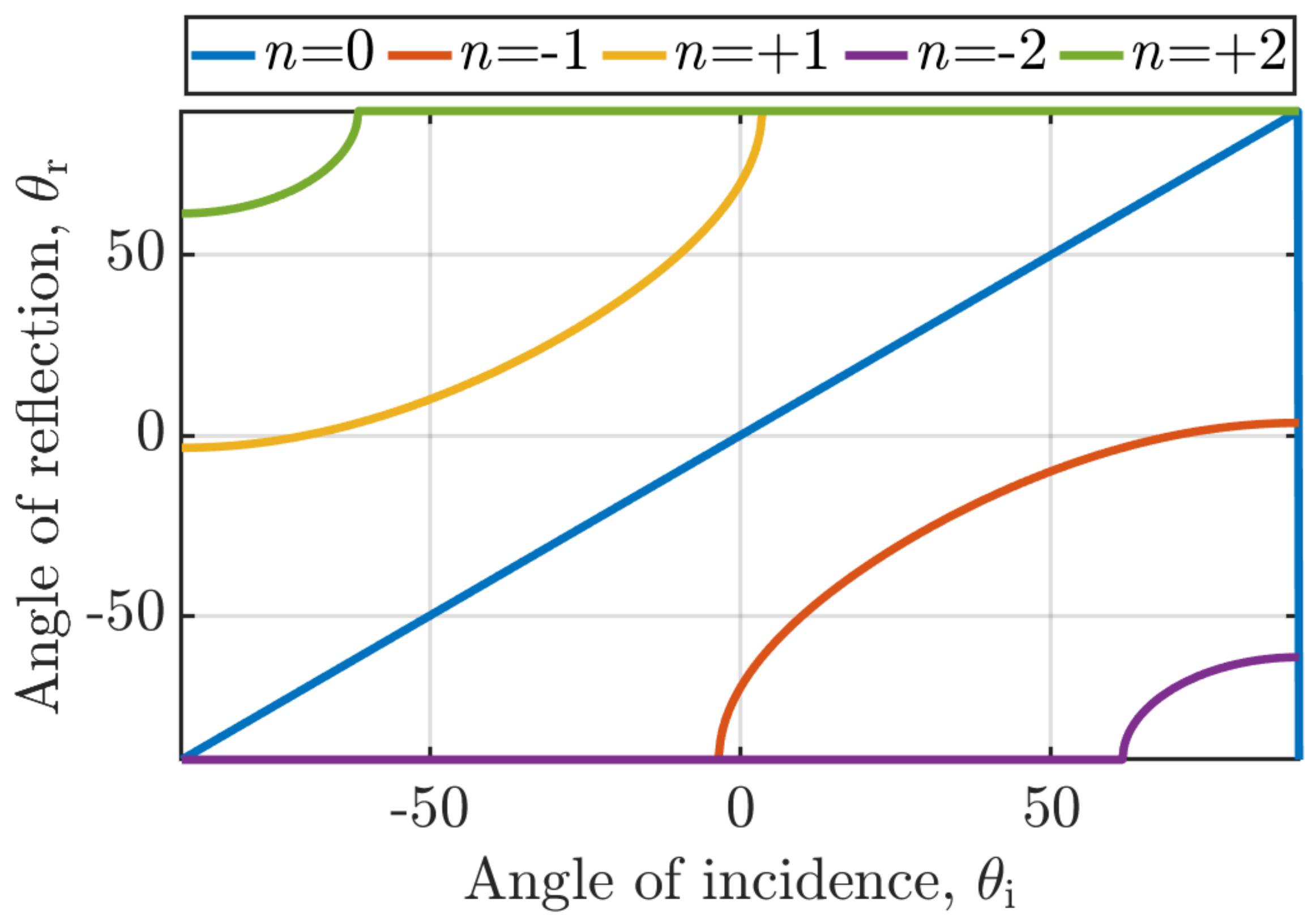}
        \caption{Propagation angle for different Floquet modes with respect to the incident angle. This figure is made using   Eq.~\ref{Eq:prop_direction} when the incidence angle is swept.}
        \label{fig:mode prop angle}
\end{figure}

A very interesting property of anomalous reflectors which is often left unstudied is the angular response, i.e., performance of the structure for various incident angles $\theta_{\rm i}$, which can be different from the design angle of incidence. Here, we consider angular response for periodical arrays formed by repeated supercells consisting of 6 unit cells with patches printed on a grounded substrate and assume the periodic boundary condition for this analysis.  To distinguish between the illumination angle and the incidence angle for which the surface was designed, we denote this design incidence angle by  $\theta_{\mathrm{id}}$. Worth to note that $\theta_{\mathrm{id}}$ together with the required reflection angle defines the period of the structure operating as an anomalous reflector for these angles. The angular response is studied by sweeping  the incident angle $\theta_{\mathrm{i}}$ for a fixed structure, designed for the angle  $\theta_{\mathrm{id}}$. The number of propagating Floquet modes existing in the system is defined by the incident angle $\theta_{\rm i}$, the period of the structure $D$, and the frequency $f$ (see Eq.~\eqref{Eq:tan_wavenumber}). The condition for the mode propagation can be written as follows:
\begin{equation}
\label{Eq:prop_modes}
 k_{\mathrm{t}n}< k_{0}\rightarrow \frac{2\pi}{D}|n|<\frac{2\pi}{\lambda}\rightarrow|n|<\frac{D}{\lambda},
\end{equation}
and their propagation directions can be calculated as~\cite{diaz2021macroscopic, diaz2021angular}:
\begin{equation}
\label{Eq:prop_direction}
\theta_{\mathrm{t}n}=\arctan (k_{\mathrm{t}n}/k_{\mathrm{n}n}),
\end{equation}
where $k_{\mathrm{n}n}$ is the normal component of the wavenumber for the \emph{n}th mode, and  $k_{\mathrm{n}n}=\sqrt{k_0^2-k_{\mathrm{t}n}^2}$. If $k_0>|k_{\mathrm{t}n}|$, the normal component of the \emph{n}th wavenumber is purely real, which corresponds to a propagating mode. Otherwise, the wavenumber is imaginary, which corresponds to a surface mode that propagates along the interface. 
Figure~\ref{fig:mode prop angle} shows that for this fixed operational frequency and period of the structure, only five  propagating Floquet modes with $n \in [-2,2]$ are allowed in the system when the angle of incidence is changing. All other modes ($|n|>3$) are surface modes.

\begin{figure*}
     \centering
         \includegraphics[width=1\textwidth]{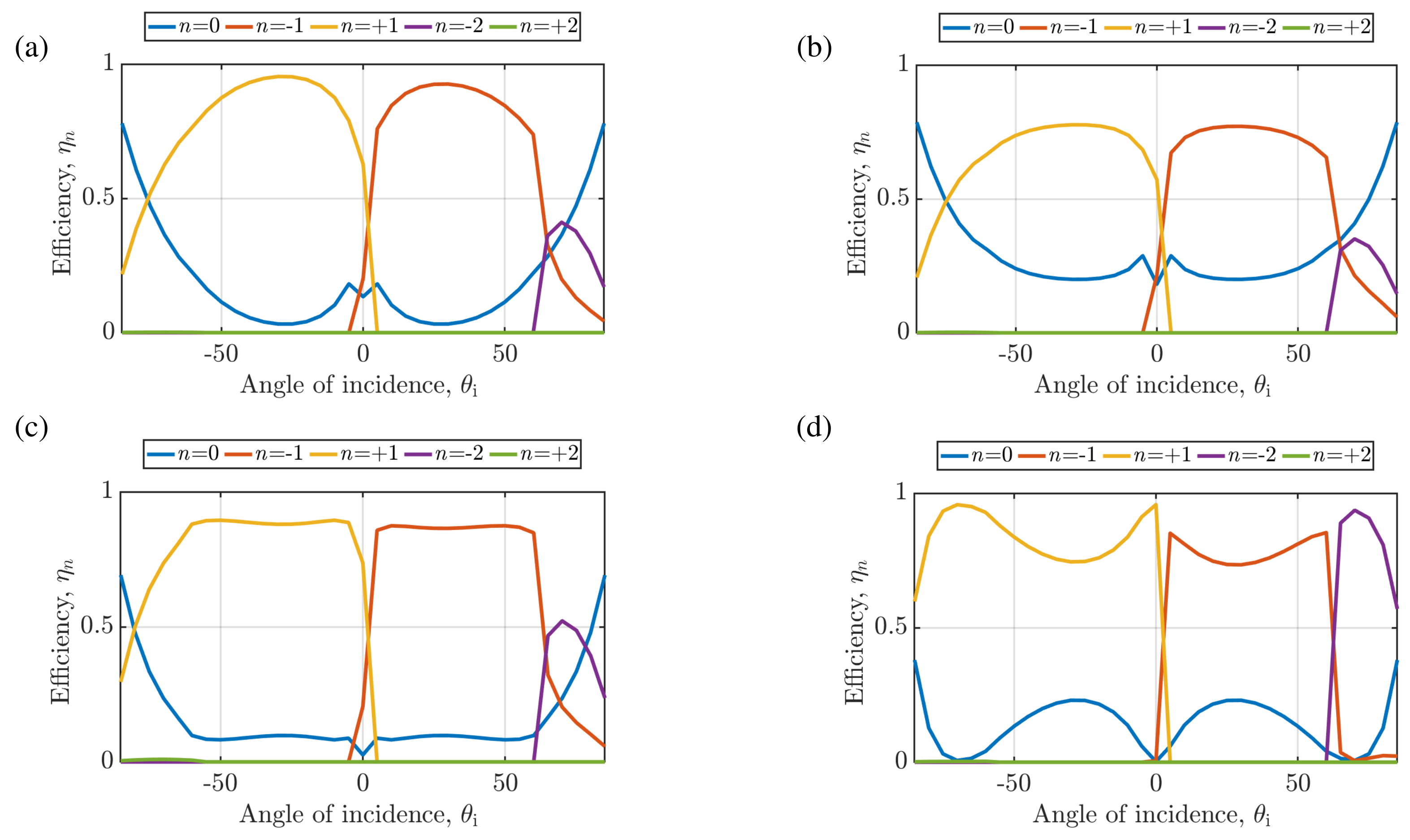}
        \caption{Power distribution among different propagating modes depending on the incident angle at frequency $8$~GHz for (a) phase gradient, (b) input impedance optimization, (c) grid impedance optimization, and (d) non-local optimization method.}
        \label{fig:spatial power}
\end{figure*}

Figure~\ref{fig:spatial power} depicts the spatial power distribution for propagating modes versus the illumination angle at $8$~GHz. At the angle $\theta_{\rm i}=0\degree$ the incident angle is equal to the design angle $\theta_{\mathrm{id}}$, therefore most of the power goes to mode $+1$, with different efficiency for each method (see Table~\ref{table:realized structure}). Based on the discussion in Refs.~\cite{diaz2021macroscopic, diaz2021angular}, for the phase-gradient case, there is a retro-reflection angle (at which all the energy is reflected back at the angle of incidence), that can be calculated as $\theta_{\rm{retro}}=\rm{arcsin}[(\sin\theta_{\rm{i}}-\sin{\theta_{\rm{r}}})/2]$ and is equal to $-28^\circ$ for the considered case. At this angle only two channels are open (see Fig.~\ref{fig:mode prop angle}), and the angle for the other channel is $-\theta_{\rm retro}$. Ideally, 100$\%$ of the power should be scattered in the  retro-reflection direction, however, discretization and the presence of losses  decrease it down to $96\%$. Therefore, the rest of the power goes to the remaining channel or gets absorbed. Due to reciprocity, the structure behaves in the same way when illuminated from direction $-\theta_{\rm retro}$.
It is important to notice that for other design methods, retro-reflection occurs at the angle $+70^\circ$, when three propagating channels are open. When the structure is illuminated from the normal direction, most of the energy couples to mode $n=+1$, where the reflection angle is $\theta_{\rm {r}}=+70\degree$. Therefore, channel $n=-1$ becomes decoupled from the other two, and when the structure is illuminated from the angle $\theta_{\rm {i}}=-70\degree$, all the energy is reflected back to the source. 

\begin{figure*}
     \centering
         \includegraphics[width=1\textwidth]{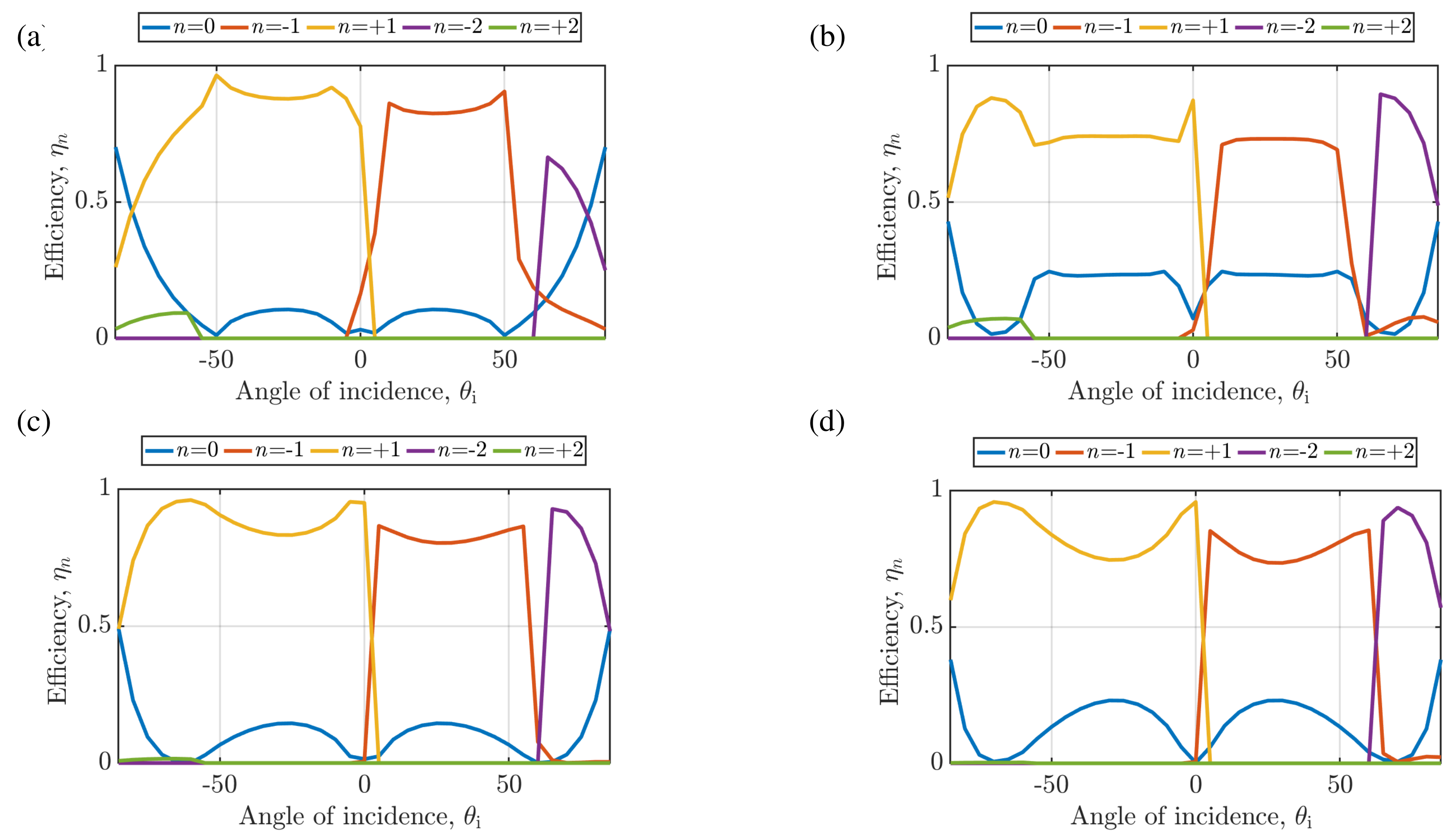}
        \caption{Power distribution among different propagating modes depending on the incident angle for the best-performance frequencies which are reported in Table~\ref{table:realized structure}, for (a) phase gradient, (b) input impedance optimization, (c) grid impedance optimization, and (d) non-local optimization method.}
        \label{fig:spatial power(best perf)}
\end{figure*}

Finally, a sweep of the incident angle reveals that the design method based on grid impedance is the solution that has the least  sensitive response (see Fig.~\ref{fig:spatial power}(c)). It means that when the incident angle changes between $-70\degree$ and $+70\degree$, the power couples primarily to the same modes, unlike for other methods. Figure~\ref{fig:spatial power(best perf)} illustrates and reports the results of the study for the best performance frequency for each method. The result is the same for the non-local optimization approach since in this case, the best performance frequency matches the design frequency.

\section{Far-field scattering from finite-size structures}
\label{section5}

In the previous analysis, we considered infinite periodical structures excited by plane waves. Here, we study far-field scattering properties of finite-size structures. 
It is possible to study metasurfaces on the grid or sheet impedance levels using the mode-matching method for calculation of induced currents \cite{hwang2012periodic,wang2020independent} and the far-field approximation for the calculation of scattered fields~\cite{diaz2021macroscopic,diaz2022integration}. To do that, the following conditions have to be met:
\begin{subequations}
\label{Eq:assumptions}
\begin{align}
|\boldsymbol{\rm{r}}|\gg \lambda, \label{Eq:assumption1}\\
|\boldsymbol{\rm{r}}|\gg L, \label{Eq:assumption2}\\
L^2/|\boldsymbol{\rm{r}}|\ll \lambda ,\label{Eq:assumption3}
\end{align}
\end{subequations}
where $|\boldsymbol{\rm{r}}|$ is the distance from the observation point to the center of the structure, and $L$ is $\rm{max}(2a,2b)$, in which $a$ and $b$ denote distances between the center of the metasurface and the edges of the structure along the $x$ and $y$-axes, respectively. Considering TE polarized incident waves ($\boldsymbol{\rm{E}}_{\rm{i}}=\rm{E_0} e^{-jk(\sin{\theta_{\rm{i}}}x+\cos{\theta_{\rm{i}}}z)}\boldsymbol{\hat y}$) and selecting the observation point in the plane of incidence ($xy$-plane), the normalized scattering pattern in spherical coordinates can be determined by the following expressions:
\begin{equation}
\label{Eq:ref_pat_anly}
F_{\rm{r}}(\theta)=\frac{1}{2\cos{\theta_{\rm{i}}}}\sum_{n} r_n (\theta_{\rm{i}})(\cos(\theta_{{\rm r}n})+\cos(\theta)){\rm sinc}(ka_{{\rm ef}\emph{n}}),
\end{equation}
\begin{equation}
\label{Eq:ref_shadow_pat_anly}
F_{\rm{sh}}(\theta)=\frac{1}{2\cos{\theta_{\rm{i}}}}(\cos(\theta)-\cos(\theta_{{\rm i}})){\rm sinc}(ka_{{\rm ef}}),
\end{equation}
where $r_n (\theta_{\rm{i}})$ are the amplitudes of excited harmonics, determined by the mode matching technique. Angle $\theta_{\rm{r}n}$ shows the reflected angle for each harmonic, and  $\rm{sinc}(x)$ is a sinc function. In addition, $a_{\rm{ef}\emph{n}}$ and $a_{\rm{ef}}$ can be represented by $a_{\rm{ef}n}=(\sin{\theta}-\sin{\theta_{\rm{r}n}})a$ and $a_{\rm{ef}n}=(\sin{\theta}-\sin{\theta_{\rm{i}}})a$, respectively. Finally, the total scattering pattern can be found as the sum  $F_{\rm{sc}}=F_{\rm{r}}+F_{\rm{sh}}$. Worth to mention that normalization is performed with respect to the maximum of the reflected field.

\begin{figure}[ht]
     \centering
         \includegraphics[width=0.45\textwidth]{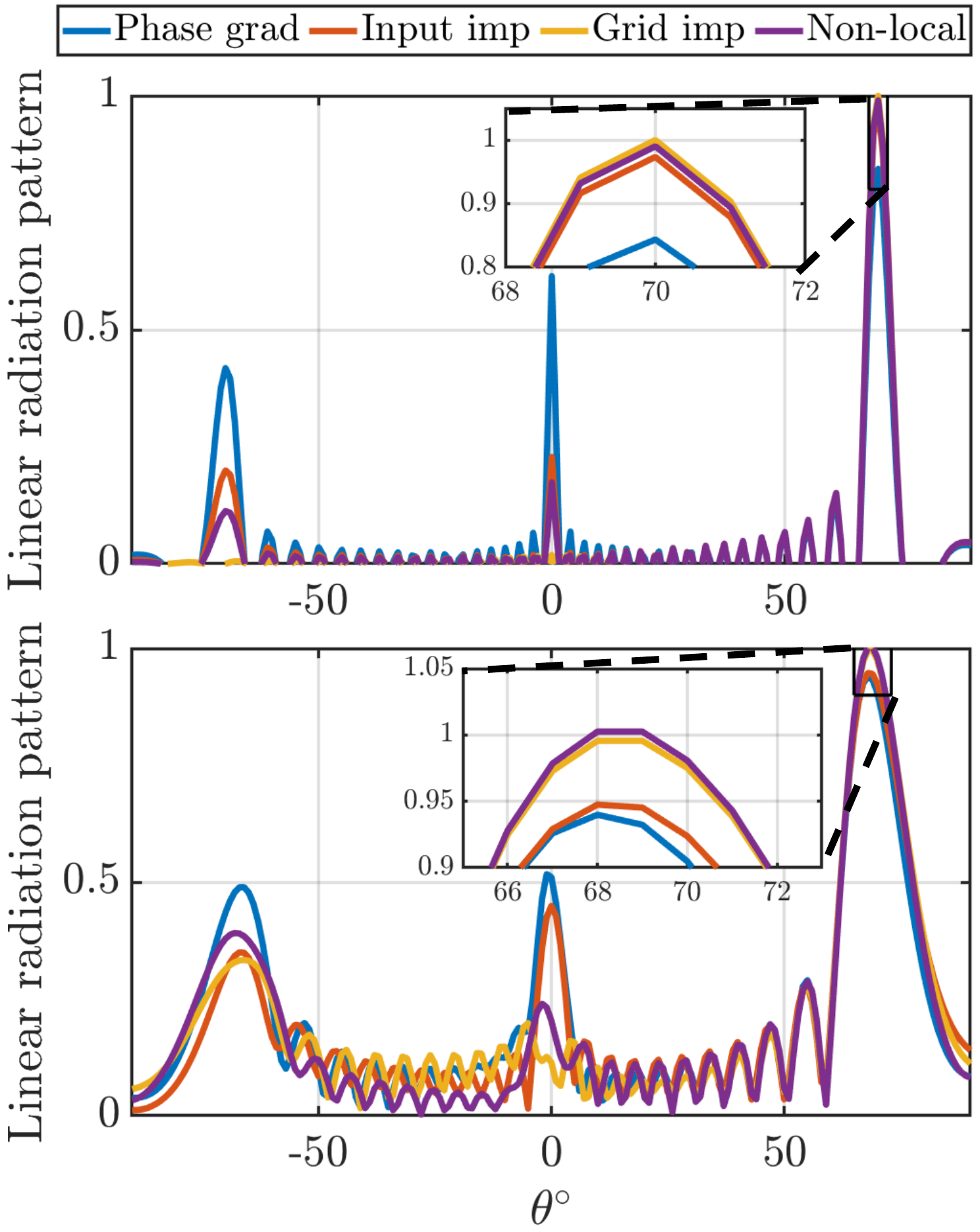}
        \caption{Normalized radiation pattern in linear scale. All patterns are normalized with respect to the main lobe amplitude of the non-local method which has the highest gain compared to the other approaches. (a) analytical pattern based on the Huygens principle, (b) full-wave simulation in CST STUDIO.}
        \label{fig:scatt dis}
\end{figure}

Alternatively to the analytical approach, one can study finite-size structures numerically using full-wave simulators. The result shown in Fig.~\ref{fig:scatt dis}(a) corresponds to the analytical solution, and in Fig.~\ref{fig:scatt dis}(b) to the full-wave simulations. In both cases, the radiation pattern is calculated at $8$~GHz for structures with the size  $11.7\lambda\times7\lambda$ in the $xy$-plane.  The discrepancy between the two radiation patterns, which becomes more significant for side lobes (SL), is caused by the neglected current distortions near the edges. Nevertheless, the general behavior is similar. Parameters of the patterns related to each method are shown in Table~\ref{table:far zone prop}.
\begin{table}[!h]
\renewcommand{\arraystretch}{1.3}
\caption{The normalized main lobe amplitude, the first side lobe amplitude in linear scale, and their directions.}
\label{table:far zone prop}
\centering
\begin{tabular}{P{0.10\textwidth}P{0.10\textwidth}P{0.10\textwidth}P{0.10\textwidth}}
\hline
 & \bfseries Main lobe Amp/angle corresponds to \emph{n}=+1.& \bfseries  SL Amp/angle corresponds to \emph{n}=0& \bfseries  SL Amp/angle corresponds to \emph{n}=-1\\
\hline\hline
Phase gradient & 0.94/68$^\circ$ & 0.52/-1$^\circ$ & 0.49/-67$^\circ$\\
\hline
Input impedance & 0.95/68$^\circ$ & 0.45/0$^\circ$ & 0.35/-67$^\circ$\\
\hline
Grid impedance & 0.99/69$^\circ$ & 0.13/-1$^\circ$ & 0.33/-67$^\circ$\\
\hline
Non-local & 1.00/69$^\circ$ & 0.24/-2$^\circ$ & 0.39/-68$^\circ$\\
\hline
\end{tabular}
\end{table}
The beam-widths for all cases are similar and close to 9$^\circ$ (see Fig.~\ref{fig:scatt dis}). Moreover, as it is shown in the inset, the maximum of the scattered field is higher for the design methods based on grid impedance and non-local solution because the power efficiency is higher in these methods compared to the phase gradient design and optimization based on the input impedance. The most important difference corresponds to the side-lobe level (SLL). As it is expected, the highest side lobes occur along the $\theta=-70$ and $\theta=0$, because there are two propagating Floquet harmonics along these directions.

\begin{figure}[ht]
     \centering
         \includegraphics[width=0.5\textwidth]{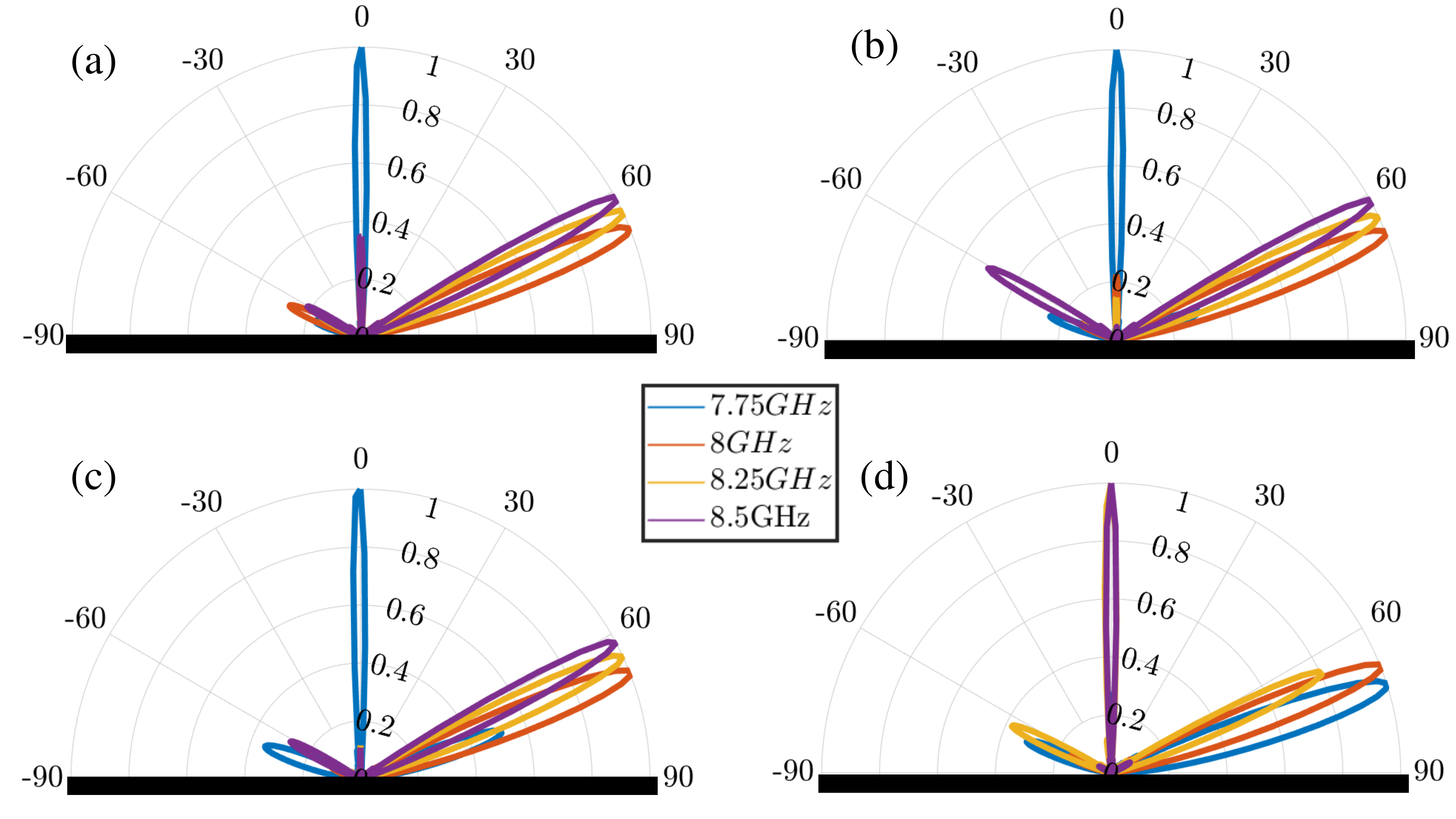}
        \caption{Frequency bandwidth patterns. At  $7.5$~GHz, which is not plotted here, there is no diffracted mode, and all the energy goes back to the specular (normal) direction. The patterns are plotted between $7.75$~GHz to $8.5$~GHz with a  $0.25$~GHz step. Designed based on (a) phase gradient, (b) input impedance optimization, (c) grid impedance optimization, (d) non-local optimization. All patterns are normalized to the main lobe amplitude for each case.}
        \label{fig:freq band}
\end{figure}

Figure~\ref{fig:freq band} shows the scattering  patterns of all the methods at different frequencies, where all the patterns are normalized to the main lobe. It is important to notice that due to the fixed period of the structure ($D=\lambda/ (\sin{\theta_{\rm{r}}}-\sin{\theta_{\rm{i}}})$), by sweeping the frequency,  the angle of reflection changes. This can be observed in Fig.~\ref{fig:freq band}. By changing the frequency, the scattered $n=+1$ Floquet harmonic scans the space from the desired reflection angle ($+70\degree$) at $8$~GHz to smaller angles at higher frequencies and larger angles at lower frequencies. Below $7.5$~GHz there are no diffraction modes, therefore we plot the scattering patterns starting from $7.75$~GHz, where most of the energy is  reflected into the normal direction (see the blue line in Fig.~\ref{fig:freq band}). The red line in the figure corresponds to the scattering  pattern at the design frequency. Eventually, the radiation patterns for $8.25$ and $8.5$~GHz are shown by yellow and purple lines, respectively.

\section{Conclusion}
\label{section6}
We have presented a comprehensive analysis of four main design methods for anomalous reflectors. In order to provide a meaningful comparison we chose design methods that can be realized within the same topology. At first, we performed an analysis of periodical infinite structures on the level of input and grid impedances. Then we proceeded to design actual implementations as supercells formed by six metal patches placed on top of a grounded dielectric substrate. Further, we analyzed the angular response of the designed metasurfaces and finally presented far-field radiation patterns of finite-size structures.

In this work, we provide a comparative summary of  the main features of previously introduced design methods as well as present an original study of a property that is frequently left unstudied: the angular response. This study can be considered referential for engineers working on reconfigurable intelligent surfaces, where similar design methods are utilized.

\section*{Acknowledgment}
This work was supported by the European Union’s Horizon 2020 research and innovation programme under the Marie Skłodowska-Curie grant agreement No 956256 (project METAWIRELESS), and the Academy of Finland (grant 345178). 

\ifCLASSOPTIONcaptionsoff
  \newpage
\fi

\bibliography{Comparison_paper}
\bibliographystyle{IEEEtran}

\end{document}